\documentclass[12pt]{iopart}

\usepackage{graphicx}
\usepackage{iopams}
\begin{document}
\title{Sedimentation of binary mixtures of like- and oppositely charged colloids:
the primitive model or effective pair potentials?}

\author{Marjolein Dijkstra$^{1}$, Jos Zwanikken$^{2}$, and Ren\'e van Roij$^{2}$}
\address{$^{1}$ Soft Condensed Matter, Debye Institute, Utrecht
University, Princetonplein 5, 3584 CC Utrecht, The Netherlands \\
$^{2}$ Institute for Theoretical Physics, Utrecht University,
Leuvenlaan 4, 3584 CE Utrecht, The Netherlands}
\date{\today}

\begin{abstract}
We study sedimentation equilibrium of low-salt suspensions of
binary mixtures of charged colloids, both by Monte Carlo
simulations of an effective colloids-only system and by
Poisson-Boltzmann theory of a colloid-ion mixture. We show that
the theoretically predicted lifting and layering effect, which
involves the entropy of the screening ions and a spontaneous
macroscopic electric field [J. Zwanikken and R. van Roij,
Europhys. Lett. {\bf 71}, 480 (2005)], can also be understood on
the basis of an effective colloid-only system with pairwise
screened-Coulomb interactions. We consider, by theory and by
simulation, both repelling like-charged colloids and attracting
oppositely charged colloids, and we find a re-entrant lifting and
layering phenomenon when the charge ratio of the colloids varies
from large positive through zero to large negative values.
\end{abstract}
\maketitle

\section{Introduction}
Suspensions of charged colloids are multi-component mixtures of
mesoscopic (colloidal) particles and microscopic cations, anions,
and solvent molecules. Due to the large size- and charge asymmetry
between the colloids on the one hand and the microscopic species
on the other, it is often difficult to treat all species on an
equal footing in descriptions of e.g. the thermodynamic
equilibrium properties of these systems. Given the asymmetry it is
natural, and often practical, to view colloidal suspensions as
"colloids-only" systems described by effective colloidal
interactions in which the presence of the microscopic particles
appear only through medium properties such as the dielectric
constant $\epsilon$ and the Debye screening length $\kappa^{-1}$.
For bulk systems in equilibrium one can in fact prove (by formally
integrating out all the microscopic degrees of freedom in the
partition function in a fixed colloid configuration) that such a
course-grained description yields the exact thermodynamics; the
problem is of course to actually perform the integration
explicitly. Within linear screening theory it is possible to
integrate out the ionic degrees of freedom explicitly, and the
resulting effective colloid pair interactions are of the
screened-Coulomb form $\propto \exp(-\kappa r)/r$, with $r$ the
center-to-center distance between the colloidal pair \cite{dlvo}.
Descriptions of suspensions of charged colloids on the basis of
this effective pair potential have often been used over the years,
see e.g. Refs.\cite{freezing,leunissen} for a few examples related
to crystallisation of charged colloids, and Ref.\cite{biben1993}
for a sedimentation study.

However, in recent studies of sedimentation equilibrium of charged
colloids at extremely low salinity this effective "colloids-only"
view was {\em not} taken. Instead the experiments
\cite{albert,mircea1,mircea2,paddy}, theory
\cite{biben1994,simonin,tellez2000,lowen,vanroij,biesheuvel}, and
simulations \cite{hynninen} of sedimentation equilibrium were
essentially all analysed within the so-called primitive model, in
which the colloids and ions are treated as charged massive hard
spheres and charged massless point particles, respectively, while
the solvent is treated as a continuum with dielectric constant
$\epsilon$. These studies have shown the existence of a
spontaneously formed macroscopic electric field that lifts the
colloids to much higher altitudes than to be expected on the basis
of their mass. This electric field is the result of the
competition between entropy (favouring a homogeneous fluid),
electrostatics (favouring local neutrality), and gravitational
energy (favouring the colloids at the bottom, but not the cations
or anions because they are essentially massless) \cite{vanroij}.
At high salt concentrations local neutrality can easily be
fulfilled at low entropic cost because of the large number of
ions, and hence the competition is the usual one between colloid
entropy and gravity, giving rise to a barometric distribution at
low packing fractions. At low salinity the situation is more
complicated, since fulfilling local neutrality would imply, when
the colloids are all close to the bottom because of gravity, that
a large fraction of the counterions must also be close to the
bottom. Such a state has a low ion entropy, because the ion
distribution is far from homogeneous. It turns out, according to
theories of e.g. Refs.\cite{biben1994,vanroij}, that the state
with the lowest grand potential spontaneously sacrifices local
neutrality at the bottom and at the meniscus, such that an
electric field is generated that lifts the colloids upward at the
expense of gravitational energy, such that the colloids and the
counterions can be rather homogeneously distributed over the
available volume, thereby increasing their entropy. For colloids
with buoyant mass $m$ and charge $Ze$ the generated electric field
strength was shown to be $E\backsimeq mg/Ze$ at low enough salt
concentrations, such that the upward electric force $ZeE$ and the
downward gravitational force $mg$ essentially cancel each other,
leading to almost homogeneously distributed colloids. Here $g$ is
the gravitational acceleration, and $e$ the proton charge. This
lift mechanism involving a macroscopic electric field was
confirmed by primitive model simulations of colloids and explicit
ions \cite{hynninen}, and by a direct measurement of the potential
difference between the bottom and the meniscus \cite{mircea2}.
Moreover, quantitative agreement was found between the measured
density profiles and the theoretically calculated ones in
Ref.\cite{paddy}, while the measured profiles of
Refs.\cite{mircea1,mircea2} agree with the theory of
Ref.\cite{biesheuvel} which involves the electric field mechanism
combined with charge regularisation.

It is at first sight not at all obvious that these phenomena,
which involve charge separation and macroscopic electric fields,
can be described within the framework of the effective
colloids-only picture. After all, in this picture the presence of
anions and cations is only accounted for in the screening length
$\kappa^{-1}$ that determines the range of the effective pairwise
interactions. However, the theoretical study of Ref.\cite{vanroij}
showed that hydrostatic equilibrium between the external gravity
field and the internal (osmotic) pressure of the effective
colloids-only system (described by the simple and crude Donnan
equation of state in Ref.\cite{vanroij}), actually predicts the
existence of the electric field and the lift effect at low
salinity, despite the underlying local density (and hence local
neutrality) approximation. The predicted density profile is linear
with height \cite{vanroij}, not unlike those predicted by a
density functional treatment of an effective colloids-only system
with screened-Coulomb interactions \cite{biben1993}. Recently
other accurate osmotic equations of state were employed and
revealed the lift effect \cite{beloni,aldemar}, and in fact
Ref.\cite{aldemar} shows that the colloid density profile that was
experimentally measured and quantitatively described in terms of
the primitive model in Ref.\cite{paddy}, can {\em also} be
described quantitatively  by a colloids-only theory with
screened-Coulomb interactions. The inconsistency of a theory based
on local charge neutrality that predicts an electric field without
explaining its sources does, apparently,  not lead to erroneous
predictions for the density profile, at least not for monodisperse
suspensions at the parameters considered.

The question we address in this paper is whether the
"colloids-only" picture can also account for the very recently
predicted layering and segregation phenomena in equilibrium
sediments of {\em binary} mixtures of charged colloids at low
salinity \cite{esztermann,zwanikken,biesheuvel2}. These studies
are actually direct extensions of the primitive model discussed
before, but now with two (or more) colloidal components with
different charges and masses. Not only was the lifting effect
recovered in Refs.\cite{esztermann,zwanikken,biesheuvel2}, but in
addition segregation into layers was found, such that the two
colloidal species order with height according to mass-per-charge:
the colloids with the lowest mass-per-charge are found at higher
altitudes. This implies the possibility of an inversion
phenomenon, whereby sufficiently highly charged heavy colloids
float on top of lighter colloids with a lower charge
\cite{esztermann,zwanikken,biesheuvel2}. In terms of the electric
field this phenomenon can be understood, at least qualitatively,
by first considering a pure system of colloidal species with mass
$m_1$ and charge $Z_1e$, such that the electric field strength is
$E\backsimeq m_1g/Z_1e$ according to \cite{vanroij}. Adding a
trace amount of colloids with mass $m_2$ and charge $Z_2e$ does
not change the field strength $E$, and hence the upward force
$Z_2eE$ on the colloids of species 2 exceeds the downward force
$m_2g$ provided $m_2/Z_2<m_1/Z_1$. In other words, if the
mass-per-charge of the tracer species 2 is smaller than that of
species 1, the former will float on top of the latter, even if
$m_2>m_1$. In the case that both colloidal species are present in
substantial amounts, the segregation into layers was found with
electric field strengths $E_i\thickapprox m_ig/Z_ie$ in the layer
with species $i$ \cite{zwanikken}. This implies a finite, nonzero
charge density in the cross-over regime from one layer to another,
as observed in Ref.\cite{zwanikken}. The question we address in
this paper is whether hydrostatic equilibrium (based on a local
density and neutrality assumption) in a colloids-only system with
screened-Coulomb interactions can catch this layering phenomenon.

There is, in addition, a second motivation for our present study.
It stems from recent experimental progress in preparing binary
mixtures with {\em oppositely} charged colloids that do not
aggregate but form equilibrium crystals \cite{leunissen}. It is of
fundamental interest to investigate the consequence of opposite
charges in the equilibrium sediment, in particular when there is a
substantial mass difference between the two colloidal species: to
what extent does the lighter species then lift the heavier one,
and to what extent does segregation into layers take place. We
therefore present not only results for colloidal mixtures with the
same charge sign ($Z_1Z_2>0$), but also with a different sign
($Z_1Z_2<0)$. Sedimentation in this latter parameter regime has,
to the best of our knowledge, not been studied at all yet. We
compare the results of computer simulations of the effective
colloids-only system with theoretical calculations based on the
Poisson-Boltzmann theory of Ref.\cite{zwanikken}.

This paper is organised as follows. In section II we present the
model of the suspension, and discuss it viewed as either a mixture
of colloids and ions as in Ref.\cite{zwanikken} or as an effective
colloids-only system that we simulate.  In section III we present
the resulting equilibrium density profiles, and section IV is
devoted to a discussion.

\section{One model and two theories}
In this section we introduce the details and the notation of our
model of a binary mixture of charged colloids that we use in this
paper, and discuss two alternative ways of describing the
equilibrium density profiles in sedimentation equilibrium.

We denote the (fixed) colloidal charges, radii, and buoyant masses
by $Z_ie$, $a_i$, and $m_i$ for the two species $i=1,2$, and
assume that the charge is homogeneously distributed on the
colloidal surface. The solvent is a continuum with dielectric
constant $\epsilon$ at temperature $T$, and the solvent volume is
$V=AH$ with $A$ the (thermodynamically large) area in the
horizontal plane and $H$ the vertical height between the bottom of
the sample (at height $x=0$) and the meniscus (at height $x=H$).
The gravitational force points in the negative vertical direction,
such that the potential energy of a colloid of species $i$ at
height $x$ is given by $m_igx$, with $g$ the gravitational
acceleration. It is convenient to introduce the so-called
gravitational length $L_i=k_BT/m_ig$, which is the characteristic
length scale of the barometric density distribution that holds in
the dilute limit.

The density profiles and the number of colloidal particles in the
sample are denoted by $\rho_i(x)$ and $N_i$, respectively, for
species $i=1,2$. We will also use the packing fraction profile
$\eta_i(x)=(4\pi/3) a_i^3\rho_i(x)$, and we characterise the
density of the suspension by the overall packing fractions
$\bar{\eta}_i=(4\pi/3)a_i^3N_i/V=(1/H)\int_0^Hdx\eta_i(x)$. The
average height of species $i$, or its center-of-mass, is defined
as \begin{equation} \label{hi} h_i=\frac{\int_0^H dx
\rho_i(x)x}{\int_0^H dx \rho_i(x)}, \end{equation} which can be
seen as a rough indication of the nature of the colloidal profile.

The suspension is imagined to be in osmotic contact with a
reservoir of massless, monovalent cations and anions of charge
$\pm e$, with a total ion density $2\rho_s$, such that the Debye
screening length (in the reservoir) is
$\kappa^{-1}=(8\pi\lambda_B\rho_s)^{-1/2}$, where
$\lambda_B=e^2/\epsilon k_BT$ is the Bjerrum length of the medium.
Here $k_B$ denotes the Boltzmann constant.

In the present calculations we restrict attention to spheres of
equal radius, and we denote the (common) diameter of the spheres
by $2a_1=2a_2\equiv\sigma$.

\subsection{Poisson-Boltzmann theory for the colloid-ion mixture}
It was shown in Ref.\cite{zwanikken} that this system can be
described by five coupled nonlinear equations for the following
five unknown profiles: the colloid densities  $\rho_1(x)$ and
$\rho_2(x)$, the two ion densities $\rho_{\pm}(x)$ for the cations
(+) and anions (-), and the electrostatic potential $\psi(x)$ or
its dimensionless form $\phi(x)=e\psi(x)/k_BT$. Two relations are
given by the Boltzmann distributions
$\rho_{\pm}(x)=\rho_s\exp[\mp\phi(x)]$ for the ions, while the
other three relations are a Boltzmann distribution for the
colloidal densities and the Poisson equation for the potential,
{\em viz.}
\begin{eqnarray}
\rho_i(x)&=&c_i\exp\left(-\frac{x}{L_i}-Z_i\phi(x)\right)\,\,\,\,(i=1,2);\nonumber\\
\phi''(x)&=&\kappa^2\sinh\phi(x)-4\pi\lambda_B\Big(Z_1\rho_1(x)+Z_2\rho_2(x)\Big),\label{pb}
\end{eqnarray}
where a prime denotes a derivative with respect to $x$, and where
$c_i$ represent normalisation constants such that $A\int_0^H
dx\rho_i(x)=N_i$. We impose boundary conditions
$\phi'(0)=\phi'(H)=0$, which corresponds to a globally charge
neutral system without any external electric field. This set of
five equations follows directly from the minimum condition on a
grand potential functional that only involves ideal-gas
contributions for all four species, a potential energy of the
colloids due to their mass, and a Coulomb energy treated at a
mean-field level \cite{zwanikken}. It is straightforward to solve
this set numerically on an $x$-grid, for details see
Ref.\cite{zwanikken}. From the solution for $\phi(x)$ the
magnitude of the electric field follows as $(k_BT/e)\phi'(x)$.

We note that these five equations are independent of the hard-core
radii $a_i$ of the colloids, i.e. the short-range part of the
(direct) correlations are not taken into account. As a consequence
this level of approximation is not capable of describing packing
effects properly. The radii $a_i$ are therefore only used to
convert $\rho_i(x)$ to $\eta_i(x)$. Despite this shortcoming the
present theory agrees quantitatively with the primitive model
simulations of Refs.\cite{esztermann,cuetos}.

\subsection{The colloids-only theory}
The Poisson-Boltzmann theory for the colloid-ion mixture differs,
at least at first sight, substantially from a description of
charged colloids based on the idea that colloids interact with
each other through effective interactions that depend on the
solvent properties such as the dielectric constant $\epsilon$, the
Debye screening length $\kappa^{-1}$, and the temperature $T$.
Standard linear screening theory \cite{dlvo} predicts that two
colloids with charges $Z_ie$ and $Z_je$ and radii $a_i$ and $a_j$,
separated by a distance $r$, interact with the pair potential
$V_{ij}(r)$ given by
\begin{equation}
\label{vij}\frac{V_{ij}(r)}{k_BT}=Z_iZ_j\lambda_B\left(\frac{\exp(\kappa
a_i)}{1+\kappa a_i}\right)\left(\frac{\exp(\kappa a_j)}{1+\kappa
a_j}\right)\frac{\exp(-\kappa r)}{r},
\end{equation}
with the Boltzmann constant $k_B$ and the Bjerrum length
$\lambda_B=e^2/\epsilon k_BT$ as introduced above already. Here we
ignore the attractive dispersion forces. Within this effective
"colloids-only" description the presence of salt ions (with a
concentration $2\rho_s$ in the reservoir) is entirely included
through the screening length $\kappa^{-1}$. We assume throughout
that the screening length in the colloidal interactions is
independent of the colloid concentration. Ignoring the counterion
contribution to $\kappa$ will turn out to be entirely justified
for all cases we study in this paper: the ratio
$y(x)\equiv\sum_{i=1}^2 Z_i\rho_i(x)/2\rho_s$ is always such that
$|y(x)|\ll1$, i.e. the background electrolyte dominates the
screening. Typically we find $|y(x)|\simeq0.01-0.2$, and since the
effective screening parameter can be written as
$\kappa^{-1}(1+y^2)^{-1/4}$ according to the Donnan-like theory of
e.g. Ref.\cite{donnan,bas}, provided the packing fraction is not
too high, we find with $(1+y^2)^{-1/4}\simeq 1-y^2/4$ that the
effective screening length differs from the reservoir value
$\kappa^{-1}$ by at most a percent.

The effective "colloids-only" description based on Eq.(\ref{vij})
has proved to be very successful in describing many facets of
colloid science, e.g. in the case of monodisperse suspensions it
explains freezing into fcc and bcc crystals at sufficiently high
densities \cite{freezing}, and it was recently used to describe
crystal structures of oppositely charged colloids successfully
\cite{leunissen}.

In this paper we use Eq.(\ref{vij}) in Monte Carlo simulations in
a box of dimensions $K\times K\times H$, with $K^2=A$ the
horizontal area. In all cases the vertical box height was taken to
be $H=109\sigma$, and the lateral width $K=10\sigma$. We checked
that $K$ was large enough to exclude finite size effects. We
employed  periodic boundary conditions in the horizontal
directions; in the vertical direction the system is bounded by
hard walls that exclude the center of colloids at heights $x<0$
and $x>H$.

\section{Results}
\subsection{Heavy colloids}
We first consider binary mixtures of equal-sized colloids with a
diameter $\sigma=1950$nm, a Bjerrum length $\lambda_B=10.4$nm, and
screening constant such that $\kappa\sigma=1.2$. This corresponds
to a salt concentration $\rho_s=2.4$nM in the reservoir. In
addition, we fix the colloidal charge of species 1 to $Z_1=76$,
and the gravitational length of species 1 to $L_1=2.5\sigma$.
These numerical values, which correspond to the experimental
system of Ref.\cite{paddy}, give rise to an effective contact
potential between the colloids of species 1 given by
$V_{11}(\sigma)=12k_BT$.  The mass of species 2 is taken to be
larger than that of species 1, by a factor of 2, such that
$L_2=1.25\sigma$.

\begin{figure}[h]
\begin{center}
\includegraphics[width=8cm]{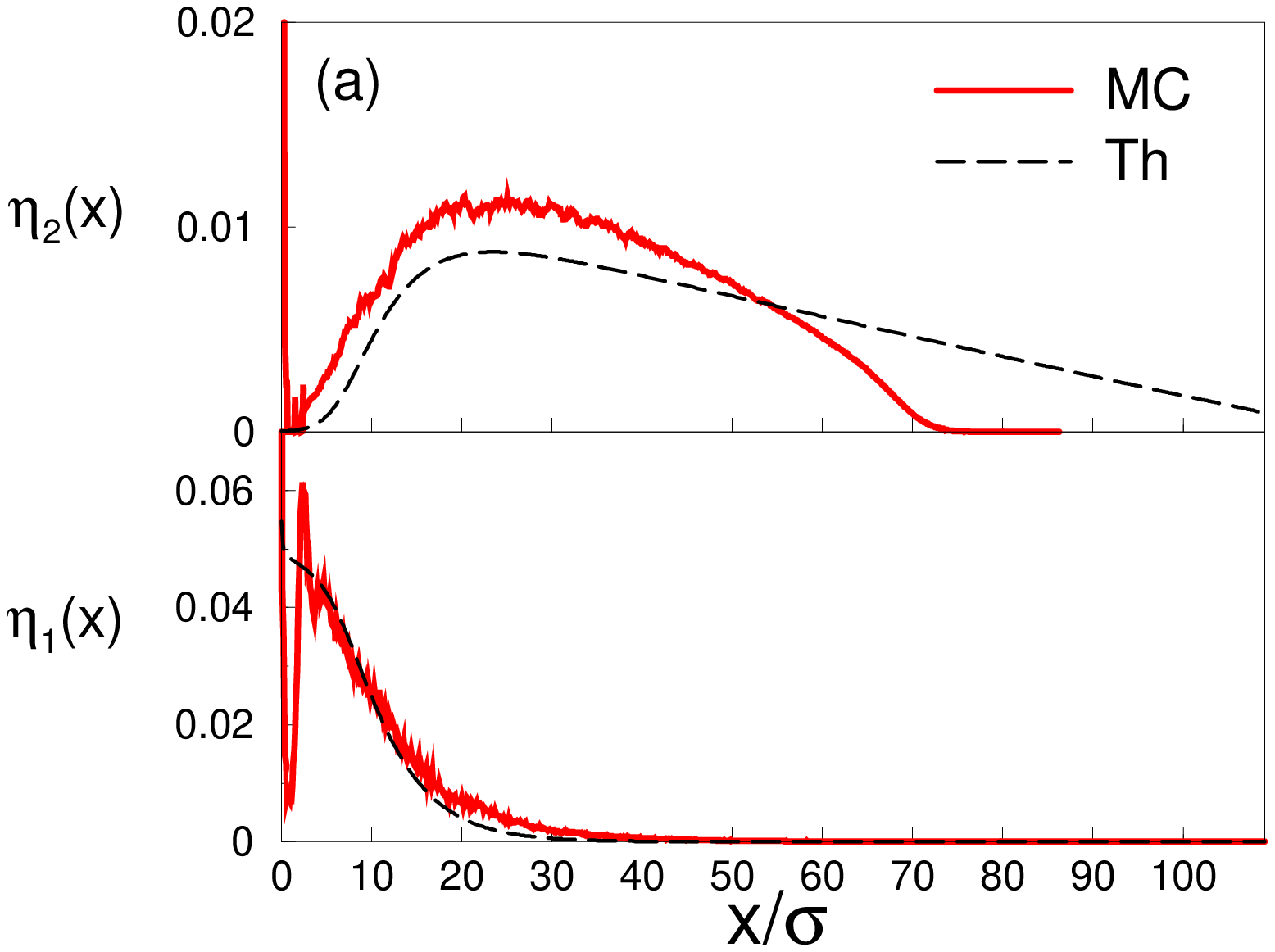}
\includegraphics[width=8cm]{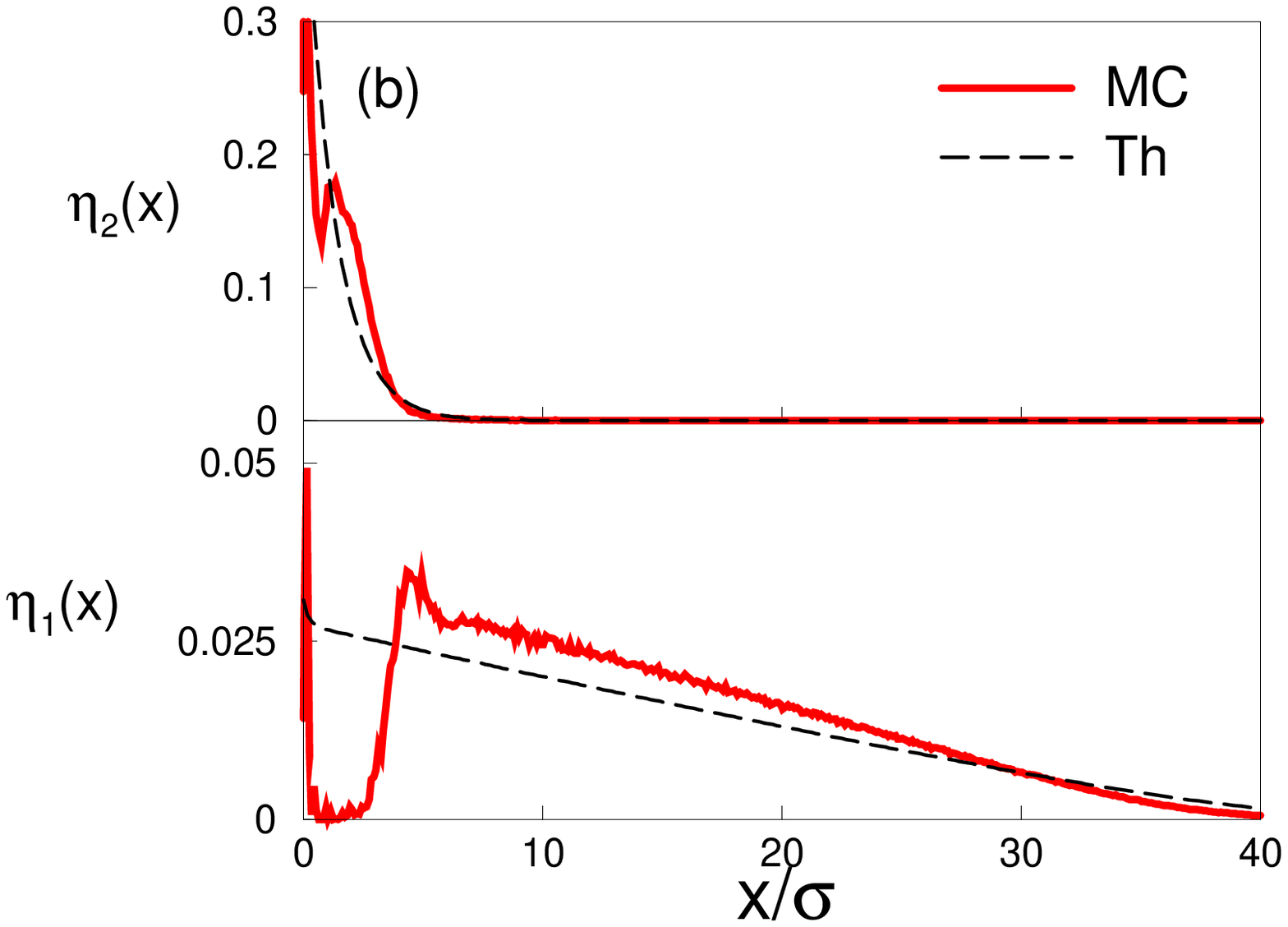}
\includegraphics[width=8cm]{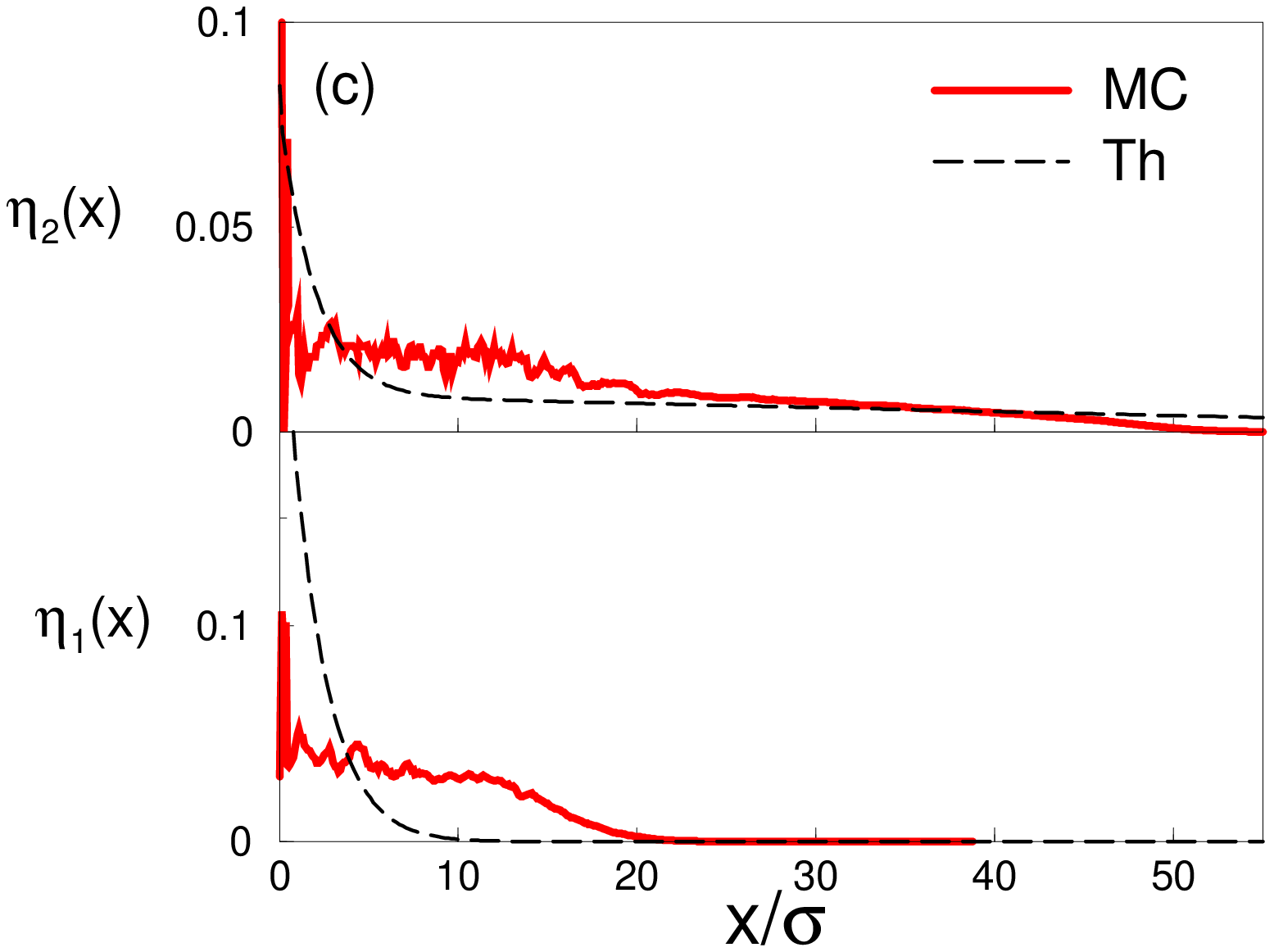}
\end{center}
\caption{Simulated (full curves) and theoretically calculated
(dashed curves) equilibrium sedimentation profiles for a binary
mixture of lighter (species 1) and heavier (species 2) colloidal
particles with charges $Z_1=76$ and various charges $Z_2$ given by
(a) $Z_2=300$, (b) $Z_2=0$, and (c) $Z_2=-300$, as a function of
altitude $x$ in a sample of height $H=109\sigma$. The common
diameter of the colloids is $\sigma=1950 ~\mbox{nm}$, the mass
ratio is $m_2/m_1=2$, the gravitational lengths are
$L_1=k_BT/m_1g=2.5\sigma$ and $L_2=1.25\sigma$, the total packing
fractions are $\bar{\eta}_1=\bar{\eta}_2=0.005$, the Bjerrum
length is $\lambda_B = 10.4 ~ \mbox{nm}$, and the screening
parameter is $\kappa\sigma=1.2$.} \label{fig1}
\end{figure}

In Fig.\ref{fig1} we plot the colloidal density profiles for three
equimolar suspensions, with average packing fractions
$\bar{\eta}_1=\bar{\eta}_2=0.005$, for three values of the charge
of species 2 given by (a) $Z_2=300$, (b) $Z_2=0$, and (3)
$Z_2=-300$. The box height is in all cases $H=109\sigma$. Case
(a), for which $Z_2L_2>Z_1L_1$, shows a layer of the heavier
species 2 floating on top of a layer of primarily species 1, both
from the theory based on
Eqs.(\ref{pb})\cite{esztermann,zwanikken}, and from the
colloids-only simulations using Eq.(\ref{vij}). The agreement
between the theory and the simulations is quantitatively
reasonable though not perfect. In particular the packing effects
shown by the simulations at length scales of the order of a few
$\sigma$ close to the hard walls of the container are not
reproduced by the theory, and the lifting effect for species 2 as
predicted by the theory is quite a bit stronger than that found in
the simulations. Nevertheless it is a striking observation that
the colloids-only picture is capable of describing the layering
phenomenon without invoking explicitly to a selfconsistent
electric field. Case (b), where $Z_2=0$ such that it represents a
mixture of charged and uncharged colloids, shows a clear lifting
effect for species 1, both from the theory and the simulations.
The theoretical prediction for species 2 is a barometric
distribution, which is bound to fail quantitatively in comparison
with the simulations given the high packing fraction up to ~0.5 at
the bottom, which causes hard-sphere like density oscillations of
species 1 close to the bottom in the simulations. Case (c) is for
oppositely charged colloids, and shows again a lifting and
layering effect, whereby a layer of the heavier species 2 floats
onto a rather dense layer at $x\lesssim15\sigma$, both in the
simulations and in the theory. Ignoring the density oscillations
on a length scale of a few $\sigma$ in this bottom layer, the
densities take the almost constant values $\eta_1(x)\simeq 0.035$
and $\eta_2(x)\simeq 0.018$, i.e. there is no large systematic
decay and the ratio of the two densities does {\em not} correspond
to $|Z_1/Z_2|$. In fact, the agreement between theory and
simulations in (c) is much less satisfactory than in the cases (a)
and (b): not only are the packing effects in the bottom layer not
captured by the theory (as before), but also the "smoothed"
simulated profiles differ substantially from the theoretically
predicted ones. We expect this to be caused by the strong
correlations that build up between attracting pairs of oppositely
charged colloids, with a contact potential of
$V_{12}(\sigma)=-47.5k_BT$, and between the repelling pairs of
species 2 with a contact potential $V_{22}(\sigma)=187k_BT$. These
correlations are not at all included in the theory. We checked
explicitly that the bottom phase in the simulations is fluid and
not crystalline.

\begin{figure}[h]
\begin{center}
\includegraphics[width=8cm]{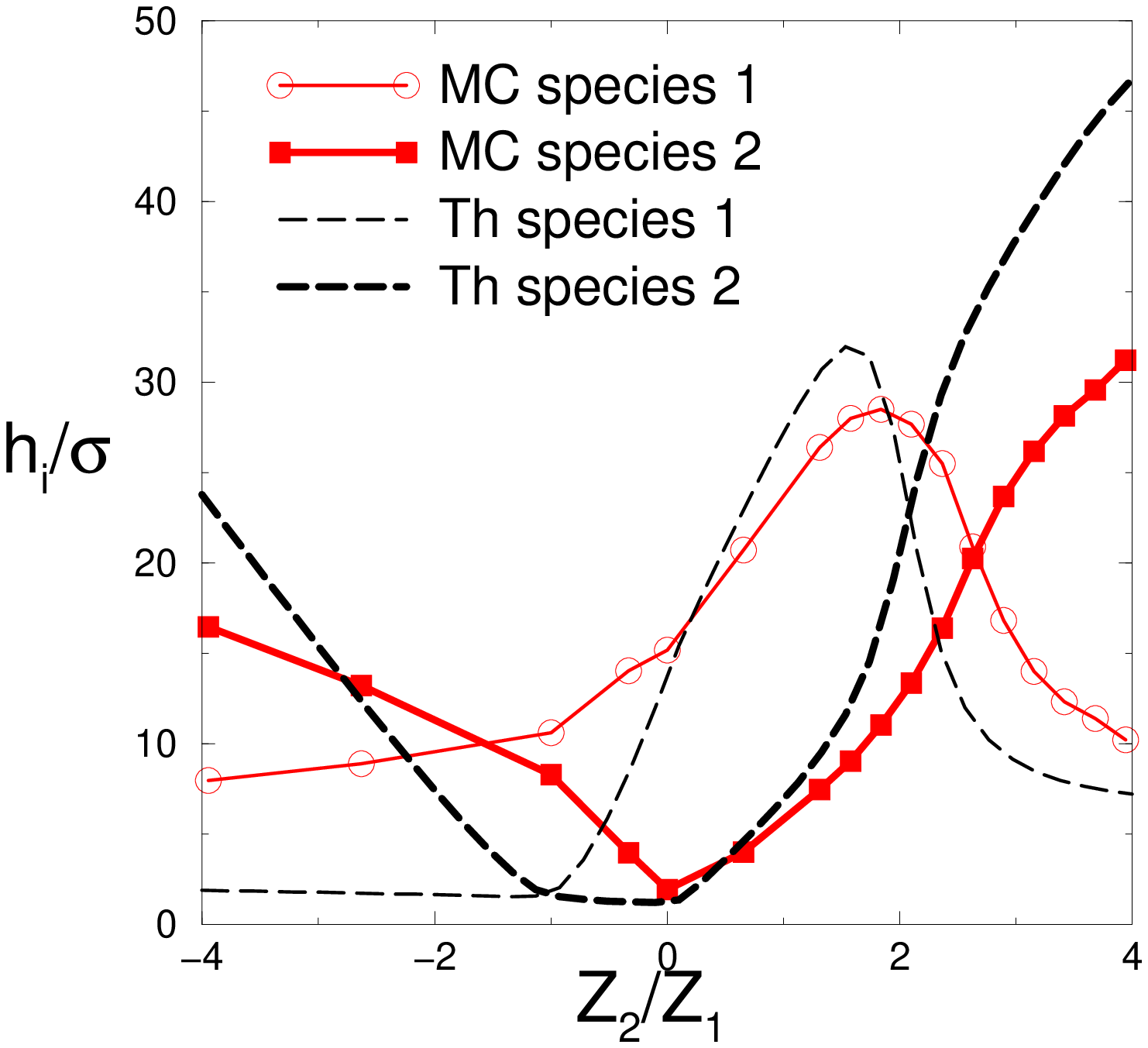}
\end{center}
\caption{Simulated (full curves) and theoretically predicted
(dashed curves) mean heights $h_i$ of light (open circles, species
1) and heavy (solid squares, species 2) colloids as a function of
the charge ratio $Z_2/Z_1$ for fixed $Z_1=76$. All other
parameters are identical to those of Fig.\ref{fig1}.} \label{fig2}
\end{figure}

An interesting observation that can be made on the basis of
Fig.\ref{fig1} is that species 2 floats onto a relatively dense
layer (with an almost constant composition) if $Z_2=\pm 300$,
while species 1 is floats onto a denser layer if $Z_2=0$. We can
quantify this re-entrant layering effect by studying the average
height $h_i$ as defined in Eq.(\ref{hi}) as a function of $Z_2$,
keeping $Z_1$ and all the other parameter fixed. The result is
shown in Fig.\ref{fig2}, and confirms the re-entrant phenomenon.
It shows gross agreement between theory and simulation for all
$Z_2/Z_1$, although the agreement is substantially better for
like-charged colloids ($Z_2>0$) than for opposite colloids
($Z_2<0$). For $Z_2/Z_1\gtrsim2$ the theory overestimates the
lifting effect for the heavier particles and underestimates it for
the lighter particles, in comparison with the simulations. For
$Z_2/Z_1\lesssim-1$ the theory predicts an essentially  barometric
distribution for the lighter species (with a lower absolute
charge), whereas the simulations still show a considerable lift
effect for species 1 in this regime, probably because they are
correlated due to the attractions of the highly charged heavier
colloids that float on top of them; this effect is poorly caught
within the present theory. Note that for the equimolar case
considered here the theory predicts $h_1=h_2$ for $Z_1L_1=Z_2L_2$
(i.e. for $Z_2/Z_1=2$ here) and for $Z_1\simeq-Z_2$ (since then
$\phi(x)\simeq 0$). On the basis of the numerical results of
Fig.\ref{fig2} these seem to be reasonable estimates for the
crossover regimes.

\begin{figure}[h]
\begin{center}
\includegraphics[width=8cm]{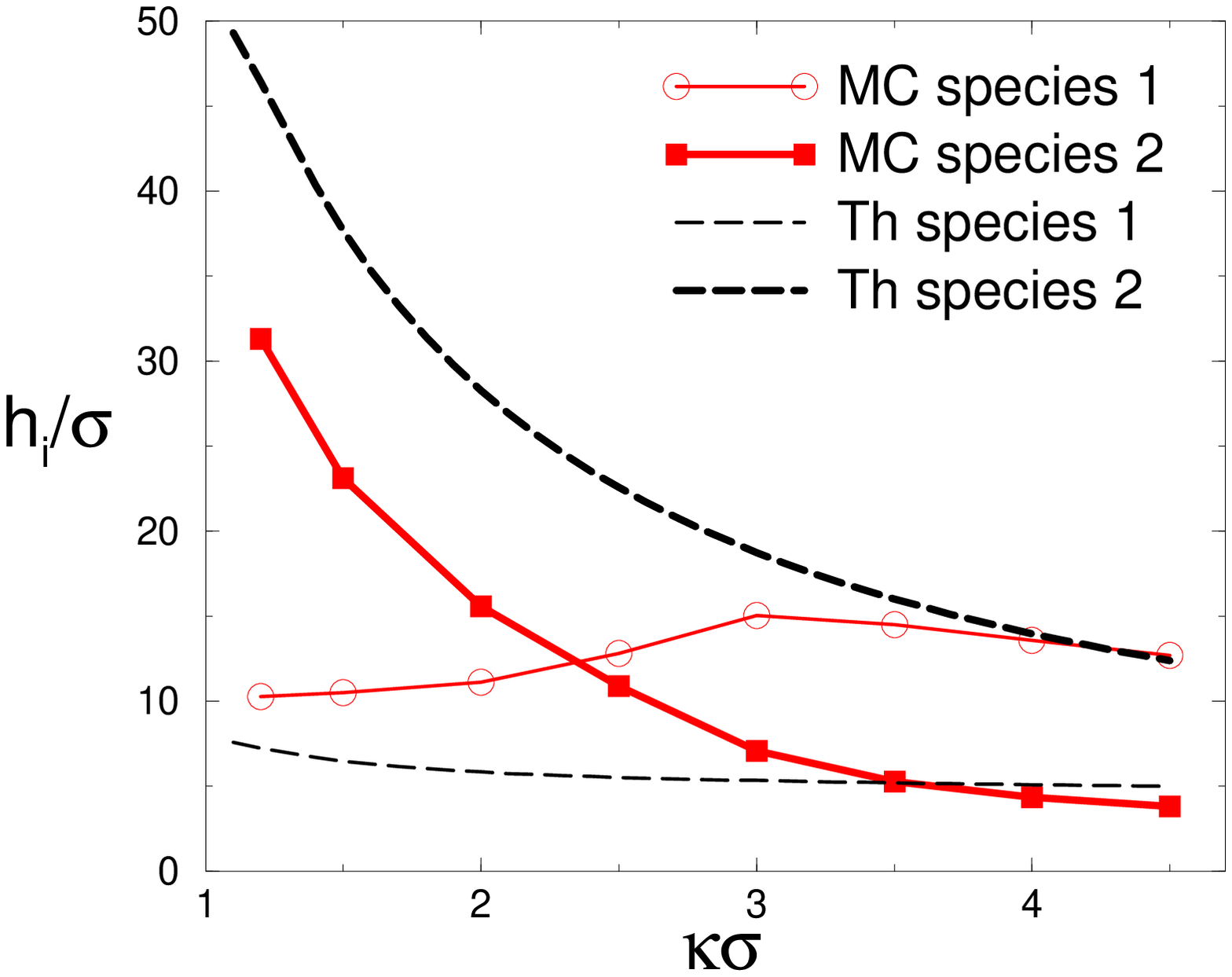}
\end{center}
\caption{Simulated (full curves) and theoretically predicted
(dashed curves) mean heights $h_i$ of light (open circles, species
1) and heavy (solid squares, species 2) colloids as a function of
the screening parameter $\kappa\sigma$  for fixed $Z_1=76$ and
$Z_2=300$ . All other parameters are identical to those of
Fig.\ref{fig1}.} \label{fig3}
\end{figure}

We also studied the effect of the screening length on the mean
height of the colloids. The result is shown, for the parameters of
Fig.\ref{fig1}(a), in Fig.\ref{fig3}. Here the theory is found to
be not at all in agreement with the simulations, except perhaps
for the decreasing mean height of the heavier species 2 with
increasing $\kappa\sigma$. The crossover where $h_1=h_2$ is found
to be at $\kappa\sigma\simeq 2.4$ in the simulations, whereas the
theory cannot locate a crossover at all for any reasonable
$\kappa\sigma$. Moreover, the theory cannot catch the phenomenon
of increasing $h_1$ with increasing $\kappa\sigma$ for
$\kappa\sigma\lesssim3$. One could argue that the theory breaks
down, not only quantitatively but even qualitatively, when
$\kappa\sigma\gtrsim 2$.

\subsection{Lighter colloids}
For the sake of comparison we repeated the study reported above to
the case where both colloidal species are ten times lighter, such
that $L_1=25\sigma$ and $L_2=12.5\sigma$. We left all the other
parameters identical to those of Figs.\ref{fig1}-\ref{fig3}.

\begin{figure}[h]
\begin{center}
\includegraphics[width=8cm]{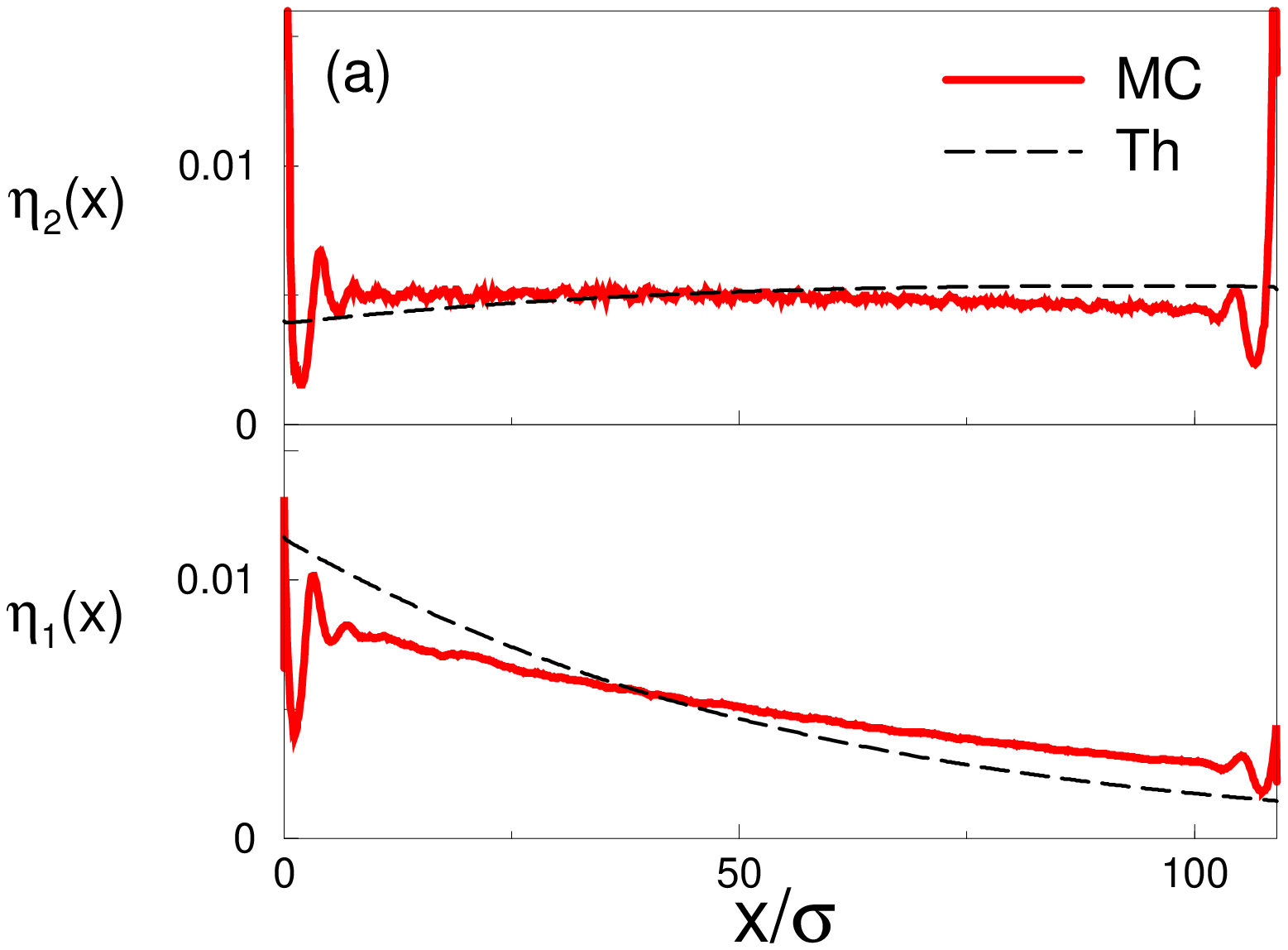}
\includegraphics[width=8cm]{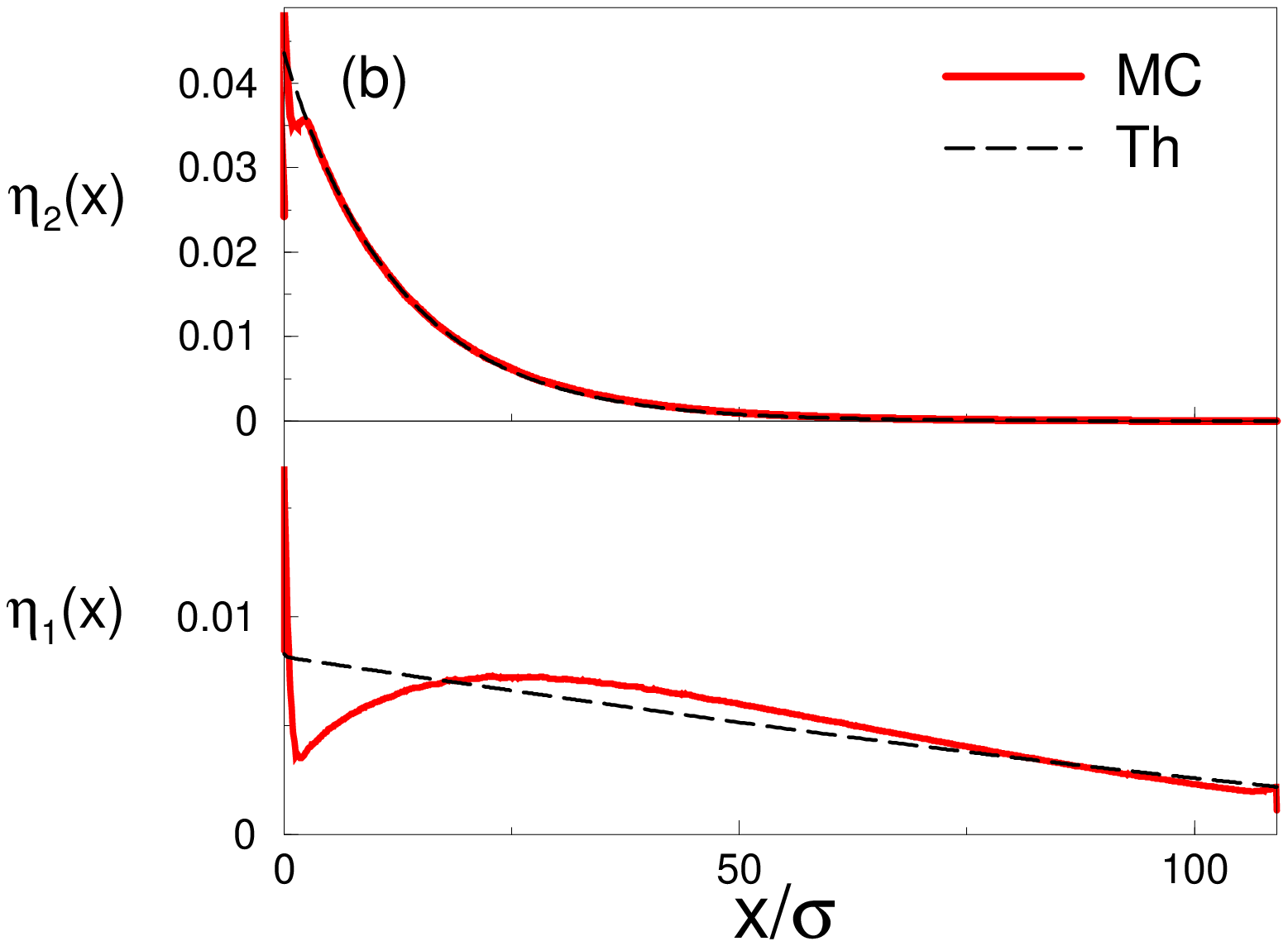}
\includegraphics[width=8cm]{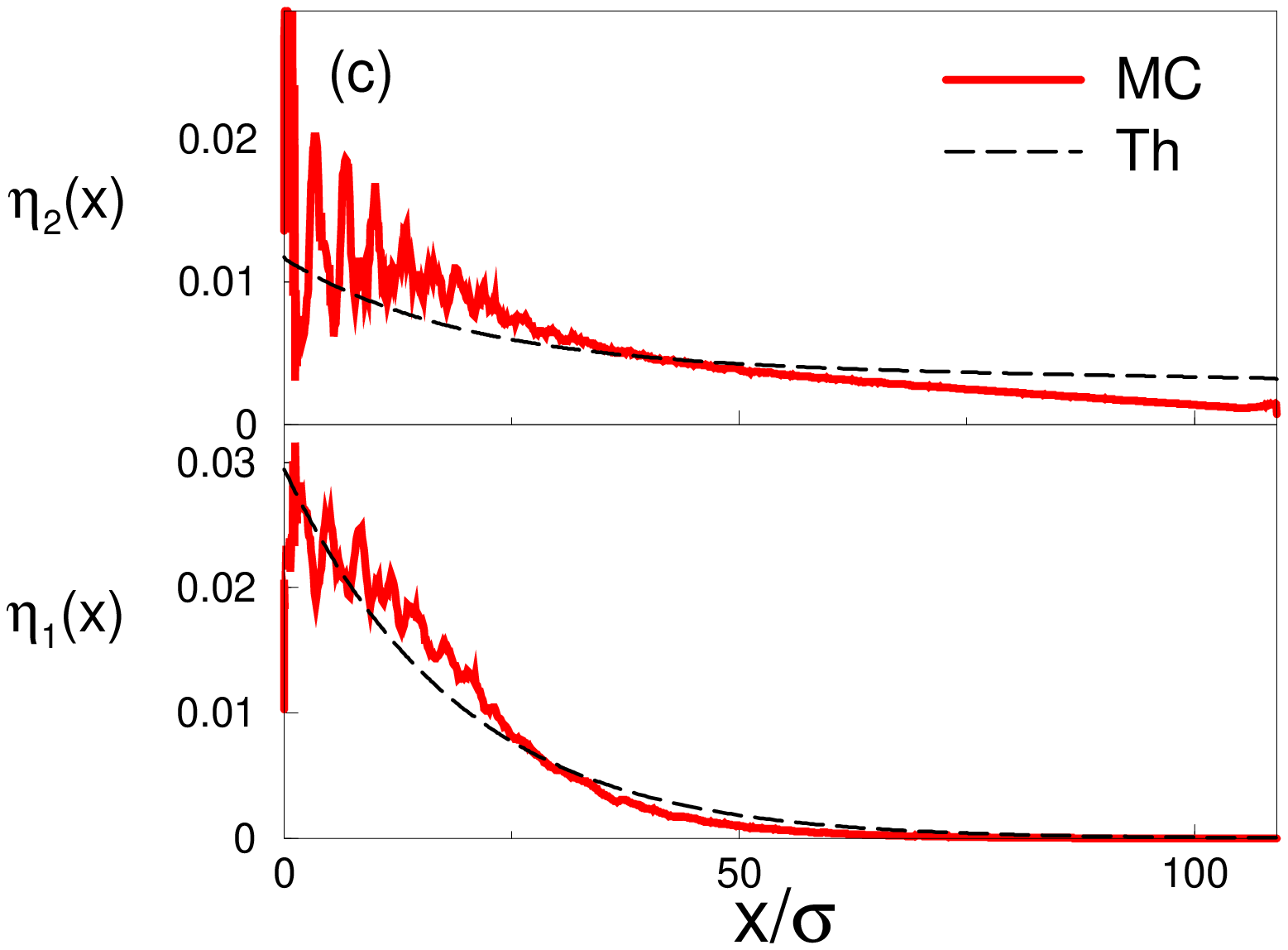}
\end{center}
\caption{Same as fig.\ref{fig1}, but for ten times lighter
colloids with gravitational lengths $L_1=25\sigma$ and
$L_2=12.5\sigma$. } \label{fig4}
\end{figure}

\begin{figure}[h]
\begin{center}
\includegraphics[width=8cm]{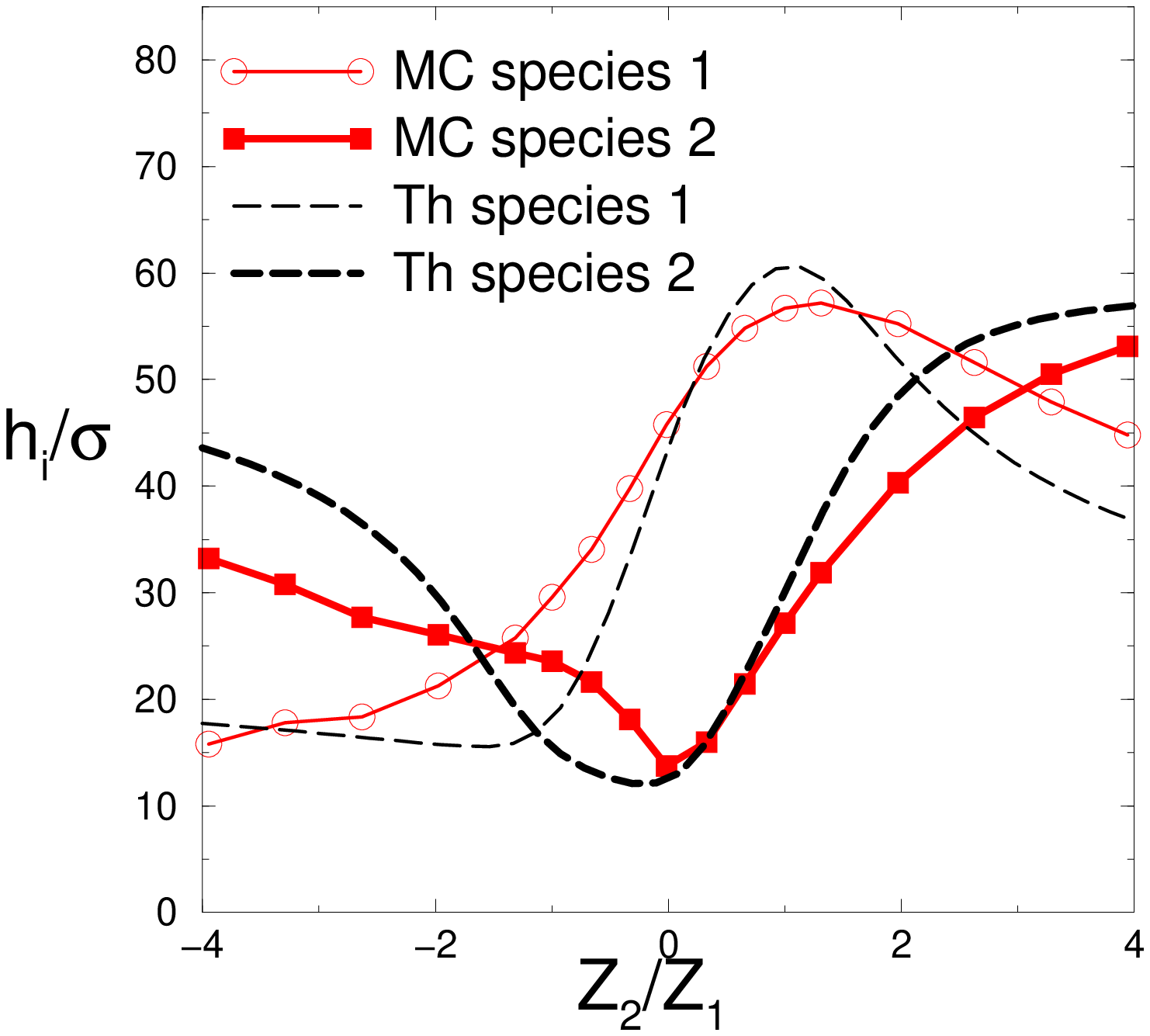}
\end{center}
\caption{Same as fig.\ref{fig2}, but for ten times lighter
colloids with gravitational lengths $L_1=25\sigma$ and
$L_2=12.5\sigma$.} \label{fig5}
\end{figure}

\begin{figure}[h]
\begin{center}
\includegraphics[width=8cm]{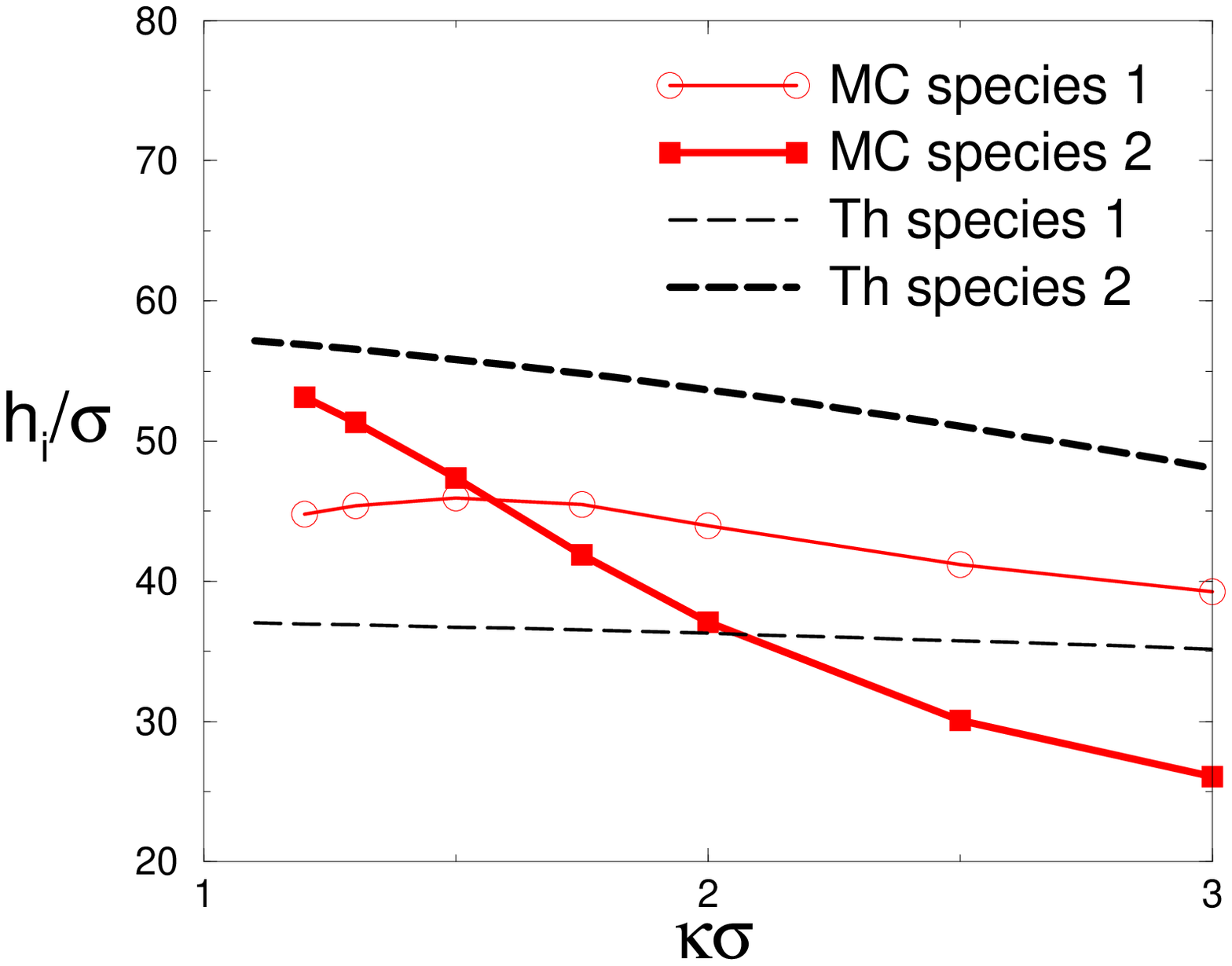}
\end{center}
\caption{Same as fig.\ref{fig3}, but for ten times lighter
colloids with gravitational lengths $L_1=25\sigma$ and
$L_2=12.5\sigma$.} \label{fig6}
\end{figure}

In Fig.\ref{fig4} we consider the three density profiles for the
three charges (a) $Z_2=300$, (b) $Z_2=0$, and (c) $Z_2=-300$, and
find reasonably good agreement between theory and simulation for
case (a) and (b), and now in fact also for (c). As in the case of
the heavier colloids the theory still fails to describe the
oscillating character of the profiles close to the bottom and the
meniscus, but in all cases it can account for a "smoothed" version
of the simulated profiles. We note that case (a) does not show a
significant layering effect, in the sense that the density
profiles of both species span the whole sample from bottom to
meniscus. This can be understood from the theory of
Ref.\cite{zwanikken} if one realises that both $Z_1L_1,Z_2L_2\gg
H$, i.e.  both species would prefer to be lifted to much higher
altitudes than the present system size $H=109\sigma$ allows.
Unlike the layering effect, the lift effect remains clearly
visible for this relatively small $H$. In Fig.\ref{fig4}(b) and
(c) the layering effect does exist, although perhaps a bit weaker
than in the case of the heavier colloids discussed above. We can
again conclude that the lifting (and layering) effect can be
accounted for within a colloids-only picture, and that there is a
reentrant phenomenon with varying $Z_2$ from 300 through 0 to -300
such that the heavier species is lifted provided that $|Z_2|$ is
large enough. This effect is quantified in Fig.\ref{fig5}, which
is the analogue of Fig.\ref{fig2} and shows the mean height $h_i$
as a function of $Z_2/Z_1$ at fixed $Z_1$. The agreement between
theory and simulation is more quantitative for these lighter
colloids, at least for like-charged colloids. This is in line with
our earlier notions that the differences between theory and
simulation are probably due to correlations, which are weaker for
lighter colloids since the system is then more homogeneous and
hence more dilute at the bottom. The analogue of Fig.\ref{fig3} is
shown in Fig.\ref{fig6}, and shows the same poor agreement between
theory and simulation as Fig.\ref{fig3}: the theory breaks down
completely for $\kappa\sigma\gtrsim 1.5$, where it cannot even
predict the correct ordering of the species with height.

\section{Conclusions}
We have studied the equilibrium sediment of binary mixtures of
charged colloids in suspension, viewed both as a colloid-ion
mixture within a Poisson-Boltzmann theory and as an effective
colloids-only system with pairwise screened-Coulomb interactions
in Monte Carlo simulations. The main conclusion is that the
layering effect, whereby colloids order with height according to
charge-per-mass as predicted by the Poisson-Boltzmann theory
\cite{esztermann,zwanikken}, can also be obtained within the
effective colloids-only system. This result is not obvious, since
the theoretical ion-colloid description invokes the ion-entropy
and a selfconsistent electric field that pushes the colloids
upwards against gravity, while the ions only occur very indirectly
in the screening constant of the effective system. Nevertheless,
these results suggest that both pictures are, at least
qualitatively, different sides of the same coin. The inconsistency
of the colloids-only system, where local electric neutrality is
assumed whereas the (Donnan) potential varies with height and thus
requires a non-vanishing local charge density at least somewhere
in the system, is apparently no serious problem for describing the
phenomenology quantitatively.

It is of interest to try to understand the layering mechanism
within the colloids-only picture qualitatively. For like-charged
colloids one may argue that the colloids with the highest charge
repel each other most strongly, such that they expand to
relatively high altitudes, leaving the lower charged (and possibly
lighter) colloids behind at lower altitudes. For oppositely
charged colloids the situation is more subtle: the attraction
between the most expanded highest charged colloids and the less
expanded lower charged colloids is such that the tendency to
layering is reduced. This effect depends, probably, sensitively on
composition and charge, and has not been investigated here in any
detail.

At a quantitative level we can conclude that the theory does not
catch the oscillatory character of the density profiles that were
found close to the bottom and meniscus in the simulations. This
effect requires a better account of the short-range correlations,
and will be addressed in future work. The theory is, all in all,
in better agreement with the simulations for repelling
like-charged colloids than for attracting oppositely charged
colloids. This can probably also be traced back to the poor level
of our theoretical treatment of the short-ranged correlations,
which are more pronounced in the presence of attractions. Given
the recent exciting new experimental developments in the study of
oppositely charged colloids exhibiting equilibrium behaviour (as
opposed to irreversible aggregation phenomena) \cite{leunissen},
there is good reason to attempt to improve the theory. Moreover,
the theory is also shown to break down completely for
$\kappa\sigma\gtrsim1.5-2$ (for the present choice of parameters),
where short-ranged correlations become more important once more.
Despite these shortcomings, our main conclusion should be that the
layering phenomenon at low salt, say at
$\kappa\sigma\lessapprox1-2$, is a real effect that can be
understood from a colloid-ion mixture perspective as well as from
a colloids-only perspective. We hope that these findings stimulate
experimental study of these phenomena.

\section{Acknowledgements}
This work is part of the research programme of the "Stichting voor
Fundamenteel Onderzoek der Materie (FOM)", which is financially
supported by the "Nederlandse organisatie voor Wetenschappelijk
Onderzoek (NWO)".

\end{document}